# Modeling hepatitis D virus kinetics during bulevirtide monotherapy: challenges and solutions

Adquate Mhlanga[1], Louis Shekhtman[1,2], Ashish Goyal[1], Elisabetta Degasperi[3], Maria Paola Anolli[3], Sara Colonia Uceda Renteria[4], Dana Sambarino[3], Marta Borghi[3], Riccardo Perbellini[3], Floriana Facchetti[3], Annapaola Callegaro[4], Scott J. Cotler[1], Pietro Lampertico[3,5,6], Harel Dahari[1*]

[1]The Program for Experimental and Theoretical Modeling, Division of Hepatology, Department of Medicine, Stritch School of Medicine, Loyola University Chicago, Maywood, Illinois, USA; [2]Department of Information Science, Bar-Ilan University, Israel;[3]Division of Gastroenterology and Hepatology, Foundation IRCCS Ca' Granda Ospedale Maggiore Policlinico, Milan, Italy; [4] Microbiology and Virology Unit, Foundation IRCCS Ca' Granda Ospedale Maggiore Policlinico, Milan, Italy; ;[5] CRC "A. M. and A. Migliavacca" Center for Liver Disease, Department of Pathophysiology and Transplantation, University of Milan, Milan, Italy [6] D-SOLVE consortium, an EU Horizon Europe funded project (No 101057917)

*Corresponding author.

[1]The Program for Experimental & Theoretical Modeling, Division of Hepatology
Stritch School of Medicine , Loyola University Chicago
2160 S. First Ave., Maywood, IL, USA 60153;
(Tel)+1-708-216-4682;
Email: hdahari@luc.edu (H. Dahari)

**Financial support:** This work was supported in part by a grant from "Ricerca Corrente RC2021/105-01", Italian Ministry of Health, and by NIH grants R01AI144112 and R01AI146917. LS thanks the Binational Science Foundation grant 207745 for support.

**Conflict of interest:** Elisabetta Degasperi: Advisory Board: AbbVie; Speaking and teaching: Gilead, MSD, AbbVie. Pietro Lampertico: Advisor and speaker bureau for BMS, Roche, Gilead Sciences, GSK, MSD, Abbvie, Janssen, Arrowhead, Alnylam, Eiger, MYR Pharma, Antios, Aligos, VIR. Harel Dahari: Advisory Board for VIR. Other authors have nothing to disclose. The other authors declare no conflicts of interest that pertain to this work.

## Abstract

The entry inhibitor Bulevirtide (BLV) was recently approved in Europe for treatment of chronic hepatitis D virus (HDV) infection, which is considered the most severe viral hepatitis infection. Theory indicates that models that account for free virus and infected cells, but do not include target cell dynamics (historically called the two-equation model) are limited to predicting a monophasic viral decline for antiviral agents that act only to block viral entry/infection. We investigated herein a recently published *two-equation* type model against clinical data obtained from patients with HDV treated with BLV monotherapy for up to 96 weeks using non-linear mixed effects modelling (NLME). We found that **(i)** although the model parameters had a relative standard error (RSE) < 50% suggesting that they were 'precisely estimated', the fits failed to reproduce the non-monophasic HDV kinetic patterns observed in most patients leading to incorrect predictions of the duration of treatment needed to reach a theoretical cure boundary, defined as less than 1 virion in the entire patient extracellular body fluid. **(ii)** The model cannot explain viral breakthrough, and **(iii)** the model wrongly predicts that viral load will remain at the same level once treatment is stopped. Lastly, we showed that including target cell dynamics in the model can explain not only monophasic viral decline during treatment but also non-monophasic HDV decline patterns such as biphasic, flat-partial response and viral breakthrough. Including target cell dynamics also predicts a viral rebound once BLV is stopped as observed in clinical studies.

## Introduction

More than 40 years since its discovery, there is still no FDA-approved therapy for chronic hepatitis D virus (HDV), the most severe form of chronic viral hepatitis (Asselah and Rizzetto, 2023). Mathematical modeling of various anti-HDV treatments is ongoing in order to understand the virus and refine possible treatment approaches (Shekhtman et al., 2023). The entry-inhibitor BLV was approved in Europe in 2020 for the treatment of chronic hepatitis D (Kang and Syed, 2020). Recently, El Massoudi and colleagues (El Messaoudi and Guedj, 2025; El Messaoudi et al., 2024) proposed a model to explain HDV kinetics under BLV-based therapy and to capture post-treatment dynamics following treatment cessation. They used nonlinear mixed effects (NLME) modeling implemented in Monolix to fit their model to data from BLV-treated patients and to estimate the model parameters. Population and individual parameter estimates, together with their relative standard errors (RSEs), were reported as part of NLME analysis.

NLME modelling is a technique employed in the field of pharmacometrics for parameter estimation across the pharmaceutical industry, US regulatory agencies, and academia (Lixoft, 2024). In NLME modelling, fixed effect estimates represent the average effect of the parameter across the population being studied, while the random effects represent the variation in the effect of the parameter across individuals within the population. RSE (standard error divided by the estimated parameter value) is a commonly used measure to determine the precision of the parameter estimate in NLME. In general, lower RSE indicates greater precision in the population estimate, while higher values indicate greater uncertainty. Several platforms such as Monolix and NONMEM offer NLME functionality with the implementation of algorithms such as the first-order method, first-order conditional estimation, and stochastic approximation expectation maximization to obtain the best fits (Bauer, 2019a; Bauer, 2019b; Chan et al., 2011). Users of Monolix have established an RSE< 50% as a threshold to identify model parameters as being considered ‘*precisely estimated,’* which has become relatively standard (AHRQ, 2022; Baaz et al., 2023; Del Amo et al., 2023;

El Messaoudi et al., 2023; Faggionato et al., 2021). Here we highlight recent work demonstrating limitations of RSE-based evaluation of population parameter estimates.

In this work, we used the El Massoudi et al. (El Messaoudi et al., 2024) model within an NLME framework in Monolix to fit HDV kinetic data in a new cohort and estimated parameters for 38 patients treated with BLV monotherapy for up to 96 weeks. We find that this model has structural limitations and thus does not capture all the observed HDV kinetics in our cohort, making it difficult to reliably estimate clinically relevant quantities such as time to cure. The model is based on a previous model that was used to study three patients treated with BLV (Shekhtman et al., 2022) who had a monophasic HDV decline, leading to an assumption of near perfect blocking of infection and supporting the application of this model, which is known to predict monophasic declines (Neumann et al., 1998). While that case study had a limited number of patients, in larger cohorts one can expand upon the individual fits and utilize population approaches to determine parameter values such as through NLME modelling. Unfortunately, in the current cohort, we find that applying NLME obscures the structural limitations of the model, which only yields monophasic kinetics. To overcome this limitation, we introduce an extended model that incorporates target cell dynamics.

## Methods

### ***Patients***

Thirty-eight patients with HDV-related advanced compensated cirrhosis were treated with BLV 2 mg/day monotherapy (Wasserman et al., 2024). All patients received Tenofovir Disoproxil Fumarate or Entecavir for hepatitis B virus (HBV). Blood samples were collected at treatment initiation, weeks 4, 8, 16, 24, 32, 40, 48 and every 12 weeks thereafter. Three patients received an orthotopic liver transplantation (at weeks 48, 60, and 72 respectively) and 1 patient stopped treatment early (week 48). HDV RNA was measured using Robogene 2.0 (lower limit of quantification, LLoQ= 6 IU/mL).

### ***Mathematical model***

The mathematical model from El Messaoudi *et al*. (El Messaoudi et al., 2024), is given by the following 3 differential equations:

$$\frac{dI}{dt} = \beta(1-\eta)VT_0 - \delta I$$

$$\frac{dV}{dt} = pI - cV \qquad (Eq. 1)$$

$$\frac{dA}{dt} = s + a\delta I - c_a A$$

where $T_0$ represents a fixed number of HDV susceptible cells, *I* the number of infected cells, V the viral load in blood and, and *A* denotes the alanine aminotransferase (ALT) concentration. The virus, V, infects target cells at a rate constant $\beta$, generating productively infected cells, *I*, which produce new virions at rate *p* per infected cell. Infected cells are lost at rate $\delta$ per infected cell, and virions are assumed to be cleared from blood at a rate $c$. The entry inhibitor BLV's blocks infection with effectiveness $\eta$, with $0 \le \eta \le 1$. The constant rate of ALT production due to HDV-infected cell death is denoted by $a$, while $s$ represents ALT production due to dying cells. Parameter $c_a$ denotes ALT clearance from blood. We assumed that the target cell level remained constant during therapy at pretreatment level $T_0 = \frac{c\delta}{\beta\rho}$. Baseline ALT, $A_0$, was estimated based on the pretreatment ALT steady-state condition: $A_0 = \frac{s+a\delta I_0}{c_a}$, where $I_0$ represents the pretreatment number of HDV-infected cells, given by $I_0 = \frac{cA_0}{p}$.

### ***Parameter estimation and fitting method***

Following El Messaoudi *et al.* (El Messaoudi et al., 2024), we fixed $\beta = 10^{-7}$ ml virions$^{-1}$ day$^{-1}$, $p = 10$ virions cell$^{-1}$ day$^{-1}$, and $c = 0.24$ day$^{-1}$. The remaining parameters ($\eta, \delta, s, a, c_a, V_0$) were estimated for each patient based on the measured viral load. Data points up to and including the first HDV measurement below the lower limit of quantification (LLoQ) or target not detected (TND) were included in the fit estimate. Each data point was given equal weight in the fitting process. We simultaneously fitted HDV and ALT data using a NLME approach. In this approach, model parameters are represented as ($\theta_i = \mu e^{\phi_i}$), where ($\mu$) represents the fixed effects describing the population average or median, while ($\phi_i$) represents the random effects accounting for inter-individual variability (IIV). Model parameters were estimated using the maximum-likelihood method implemented in Monolix version 2023R1 [http://software.monolix.org]. The estimated parameter confidence intervals (CIs) were computed at a 95% level of significance with a bootstrap method using R version 4.3.2 utilizing the Rsmlx package with 100 bootstrap samples, on the same data as previously described above (Wasserman et al., 2024) and parameter estimates (**Table 1**) obtained from Monolix.

## Results

In our cohort, population parameter estimates were all suggested to be 'precisely estimated' with RSE < 50% (**Table 1)**, which remained true when accounting for the correlation of random effects. This result in our cohort replicates the findings of El Massoudi et al. (Eq. 1) in their cohort and would suggest that the model is able to fit well the observed data in our cohort.

| | **Population Estimates** | | **Inter Individual Variability** | |
|---|---|---|---|---|
| Parameter | Value [95% Confidence Interval] | RSE% | Value [95% Confidence Interval] | RSE% |
| $\eta$ | 0.86 [0.40 – 0.999] | 17.00 | 1.61 [0.58 – 13.19] | 59.0 |
| $\delta$ | 0.011 [0.01 – 0.029] | 24.80 | 0.94 [0.35 – 1.40] | 16.9 |
| $\log V_0$ | 3.76 [3.46 – 4.21] | 4.98 | 0.28 [0.18 – 0.34] | 14.5 |
| $c_a$ | 2.33 [1.80 – 3.00] | 2.60 | 0.07 [0.02 – 0.08] | 26.4 |
| $a$ | 0.143 [0.035 - 0.346] | 49.70 | 2.13 [1.33 – 2.73] | 19.9 |
| $s$ | 3.48 [2.71 – 4.44] | 2.60 | 0.06 [0.04 – 0.11] | 27.7 |

**Table 1:** Estimates of the model parameters are presented along with their RSE% and confidence intervals (CIs). In non-linear mixed effects (NLME) models the population estimate is described by the fixed effects and the inter-individual variability (IIV) by the standard deviation (SD) of the random effects.

Despite these apparently promising results from NLME, we carried out further assessments and followed the modeling workflow established by Traynard et al. (Traynard et al., 2020) to identify potential issues. First, we reviewed the diagnostic plots (**Fig. 1** and **Supplementary Figs. 1 - 4**) and quickly determined that the fits appear accurate for patients whose declines exhibited monophasic-like characteristics (**Fig. 1A-B**); However the fits fail to recover the multiple slopes for patients with a biphasic decline (**Fig. 1C-D**). **Supplementary Fig. 1** (observations vs. predictions) showed a lack of symmetry, indicating potential model fitting problems. Likewise, a visual predictive check, VPC (**Supplementary Fig. 2**) highlighted the presence of outliers. of Individual weighted residuals (IWRES) vs. time and IWRES vs. individual predictions displayed asymmetry (**Supplementary Fig. 3)** and IWRES vs. time and IWRES vs. individual predictions also exhibited asymmetry (**Supplementary Fig. 4)**. Together, these findings indicate issues with the model design even though parameters met the threshold to be considered precisely estimated. Below we further detail how these mathematical issues can give rise to misleading clinical interpretations with possible important implications for predicting time to cure.

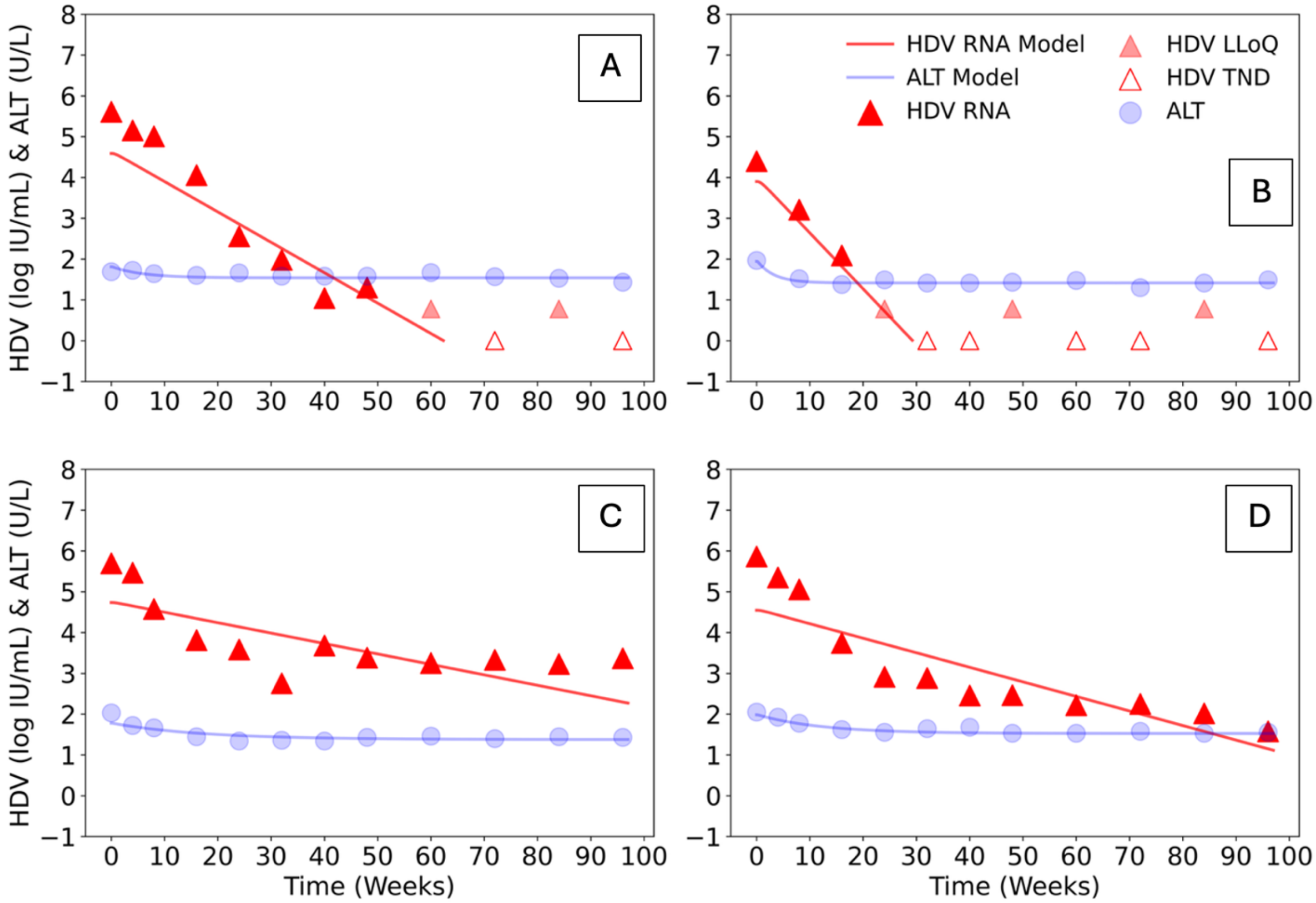

**Fig. 1**: Model calibration (curves) with individual patient HDV RNA and ALT kinetic data (symbols) during 96 weeks of BLV monotherapy was performed in Monolix and visualized using Python version 3.11. **(A)** (patient 8) and **(B)** (patient 26) show declines with monophasic-like decline characteristics, whereas **(C)** (patient 3) and **(D)** (patient 16) show clearly non-monophasic patterns, which are not explained by the model. Lower Limit of Quantification (LLoQ), Target not detected (TND).

### *Estimates of time to cure*

Poor fits based on incorrectly specified models not only lead to incorrect estimates of biological parameters, but can also have clinical implications. A key contribution of mathematical modelling is the ability to calculate potential time to cure, namely the time point when there is less than 1 virion in the entire body fluid (the cure boundary) as previously predicted for HDV under BLV monotherapy (Shekhtman et al., 2022) and done successfully for hepatitis C on treatment (i.e., in real-time) (Dahari et al., 2015; Etzion et al., 2020). For patients experiencing continued viral declines, this is typically done by extending the fitted curve for a longer time period and assessing when the cure boundary is reached. Crucially, if the estimated slope of the last phase of the kinetics is incorrect, the expected time to reach the cure boundary will be incorrect, potentially leading to crucial clinical errors. In **Fig. 2**, we demonstrate an estimated prediction of the time to cure for two patients undergoing a flat-partial response (**Fig. 2A**) and biphasic decline (**Fig. 2B**). The El Massoudi et al. model predicts a monophasic decline, leading to a much earlier estimate of the time to cure than would be expected based on the reduced rate of decline observed after around week 40, where there is a flattening of the decline in the flat-partial response patient (**Fig. 2A**) and the slower second phase in the biphasic patient (**Fig. 2B**) .

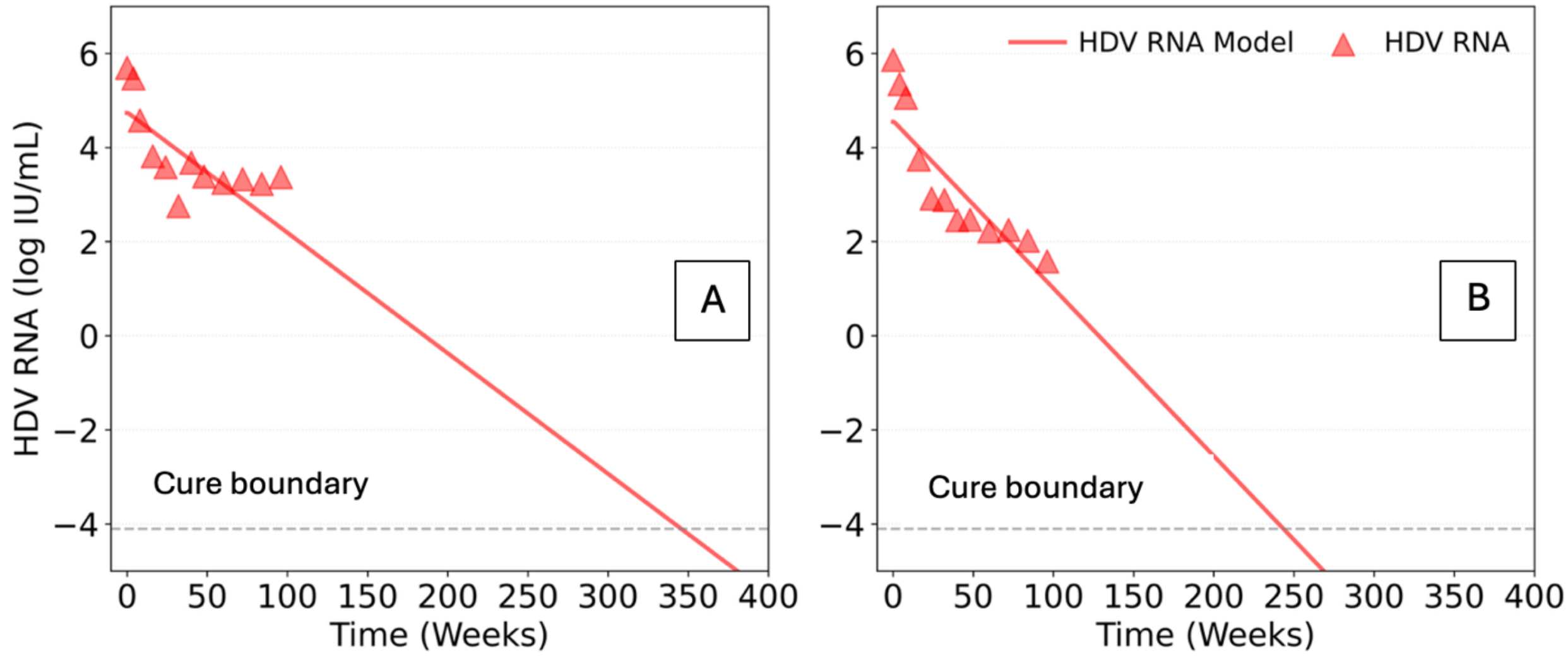

**Fig. 2**: Model fits of HDV for two representative patients. **(A)** shows the El Massoudi model fit curve for patient 3 (flat-partial response pattern), and **(B)** for patient 16 (biphasic decline pattern). Red triangles denote measured HDV levels (log IU/mL), and solid red lines denote the corresponding El Massoudi model predicted HDV curves over time (weeks). The dashed grey line represents the cure boundary.

### ***Breakthrough Kinetics Cannot be Explained by the El Massoudi Model***

Another issue we observe with applying the model of El Massoudi et al. (Eq. 1) to our cohort is that several patients experienced breakthrough kinetics. Following an initial decline, we observe a subsequent increase in the viral load in several patients rather than a continuous decline as the model would suggest. In **Fig. 3**, we demonstrate examples of such patients, showing that the fit clearly does not correspond with the observed kinetics. Notably, in the original work of El Massoudi et al. (El Messaoudi et al., 2024) we identify several such patients with breakthrough kinetics such as in **Fig. 3** of their manuscript where there is a curve that at 36 weeks is around 3 log IU/mL but at near 48 weeks increases to around 6 log IU/mL). Thus, both in our cohort and El Massoudi et al. cohort (El Messaoudi et al., 2024) a monophasic fit does not reflect breakthrough kinetics.

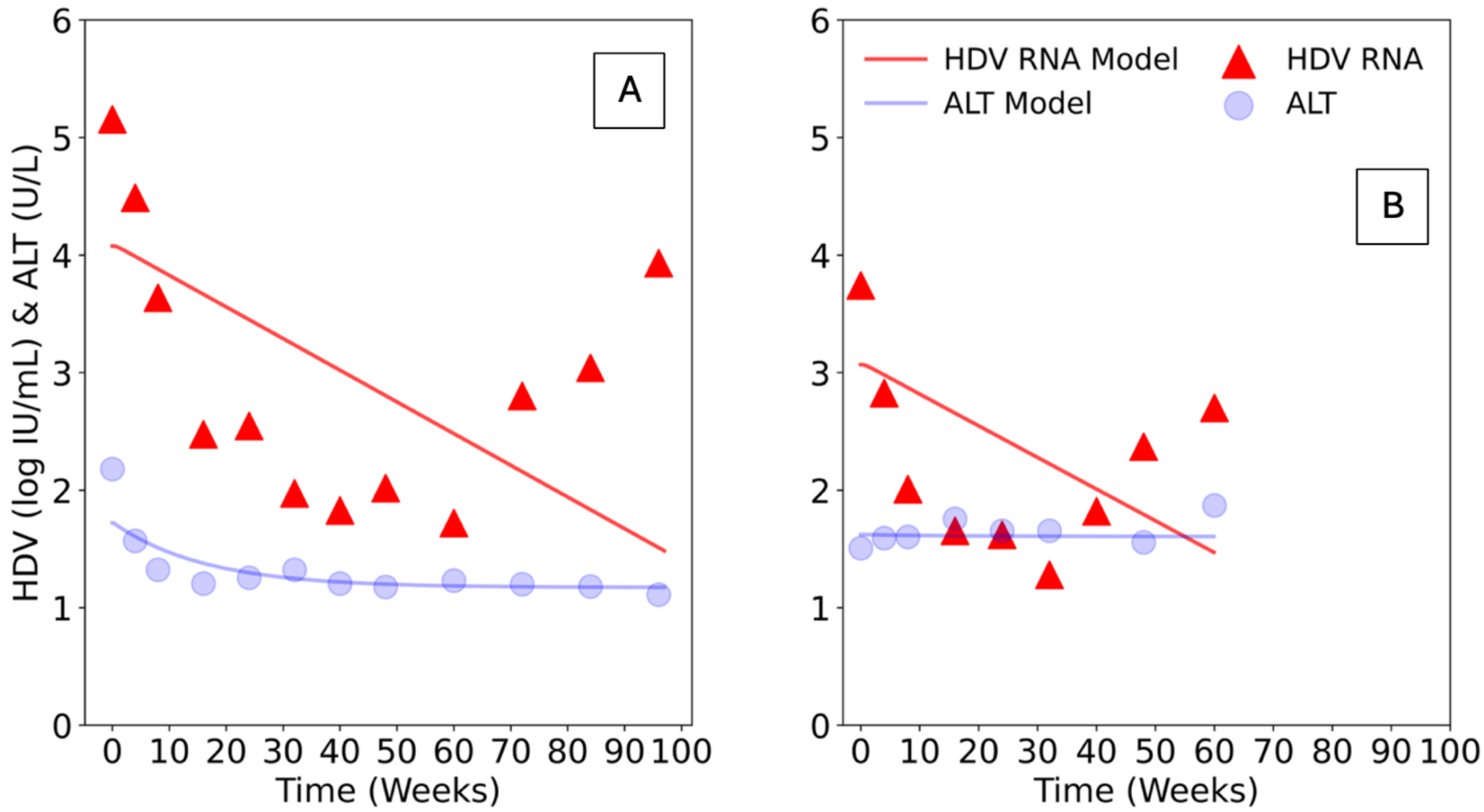

**Fig. 3**: Model fits using the 2-equation El Massoudi et al. model for patients with viral breakthrough. **(A)** Patient 1 (late breakthrough, > 48 weeks). **(B)** Patient 16 (early breakthrough, < 48 weeks). Red triangles: observed HDV (log IU/mL); red line: model predicted HDV. Blue circles: observed ALT (U/L); blue line: model predicted ALT.

***Failure to capture post-treatment dynamics***

We investigated how the El Massoudi model predicts viral kinetics after discontinuation of BLV (El Messaoudi and Guedj, 2025; El Messaoudi et al., 2024). In **Fig. 4** we show that the El Massoudi model predicts that viral load will remain steady after the completion of BLV therapy. In contrast, clinical studies show viral rebound following cessation of treatment (Deterding et al., 2025; Wedemeyer et al., 2025; Zoulim et al., 2025).

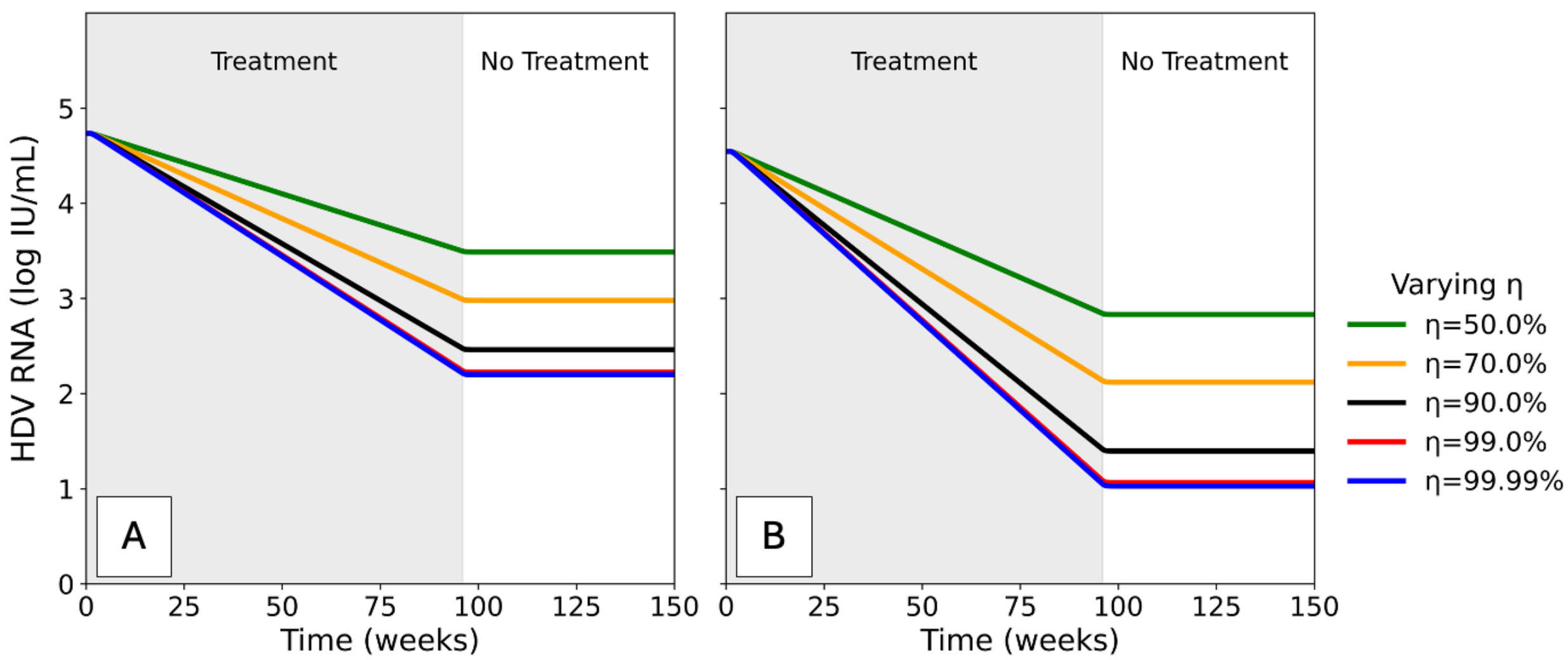

**Fig 4**. **Simulated treatment cessation, with A (patient 3) and B (patient 16)**. The shaded grey area indicates the on-treatment phase (weeks 1 - 96); the unshaded area indicates the post treatment phase. Here we show log HDV for varying $\eta$ = [99.99%, 99%, 90%, 70%, 50%]. HDV remains steady following treatment cessation in all cases.

***An extended model that includes target cell dynamics can predict all observed viral kinetic patterns during treatment and rebound post treatment cessation***

We provide herein a solution to the structural limitation of El Masoudi model (Eq. 1) by replacing the fixed number of target cells (i.e., $T_0$ in Eq .1) with a dynamic equation of target cells (Eq. 2) as follows:

$$\frac{dT}{dt} = s_T - dT - (1-\eta)\beta VT \qquad (Eq.2)$$

Where target cells (T) are assumed to be produced at constant rate $s_T$ and die with a death/loss constant rate $d$.

In Fig. 5 we show that a model that includes target cell dynamics can predict all three kinetic patterns that the El Massoudi et al. model is structurally unable to capture during BLV monotherapy including flat-partial response (**Fig. 5A**), biphasic (**Fig. 5B**), and breakthrough (**Fig. 5C**). In **Fig. 6A** we show the distinct difference between the models for a patient with a flat partial response, where the El Massoudi model predicts continued decline whereas the extended model predicts an increase. Likewise in **Fig. 6B**, for the biphasic kinetics, the extended model incorporates the second phase slope and thus leads to a much longer estimate of predicted time to cure.

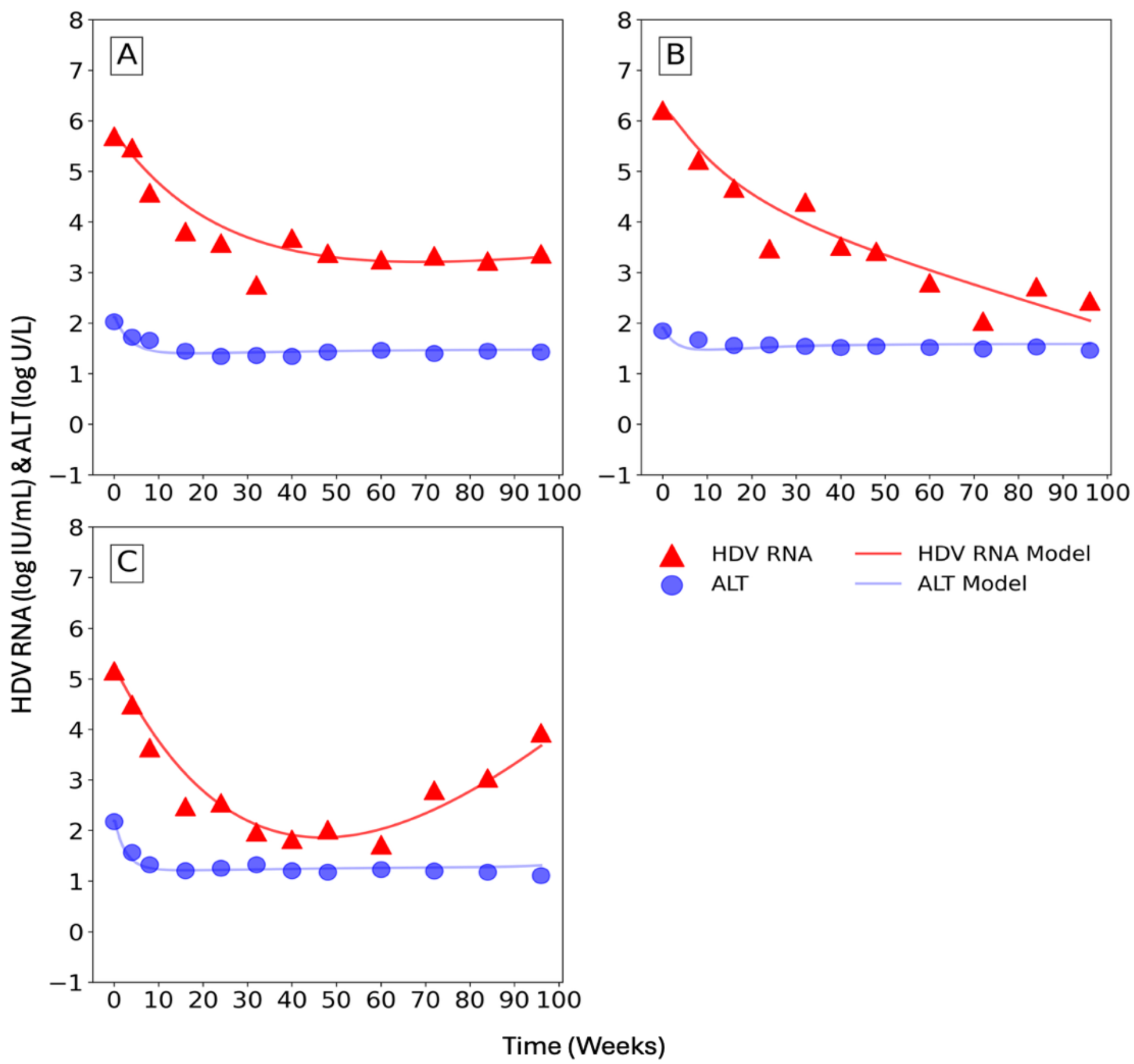

**Fig 5**. **Extended model incorporating dynamic target cells.** We show that an extended model incorporating dynamic numbers of target cells can recover **(A)** flat-partial response (patient 3), **(B)** biphasic (patient 44) and **(C**) viral breakthrough (patient 1).

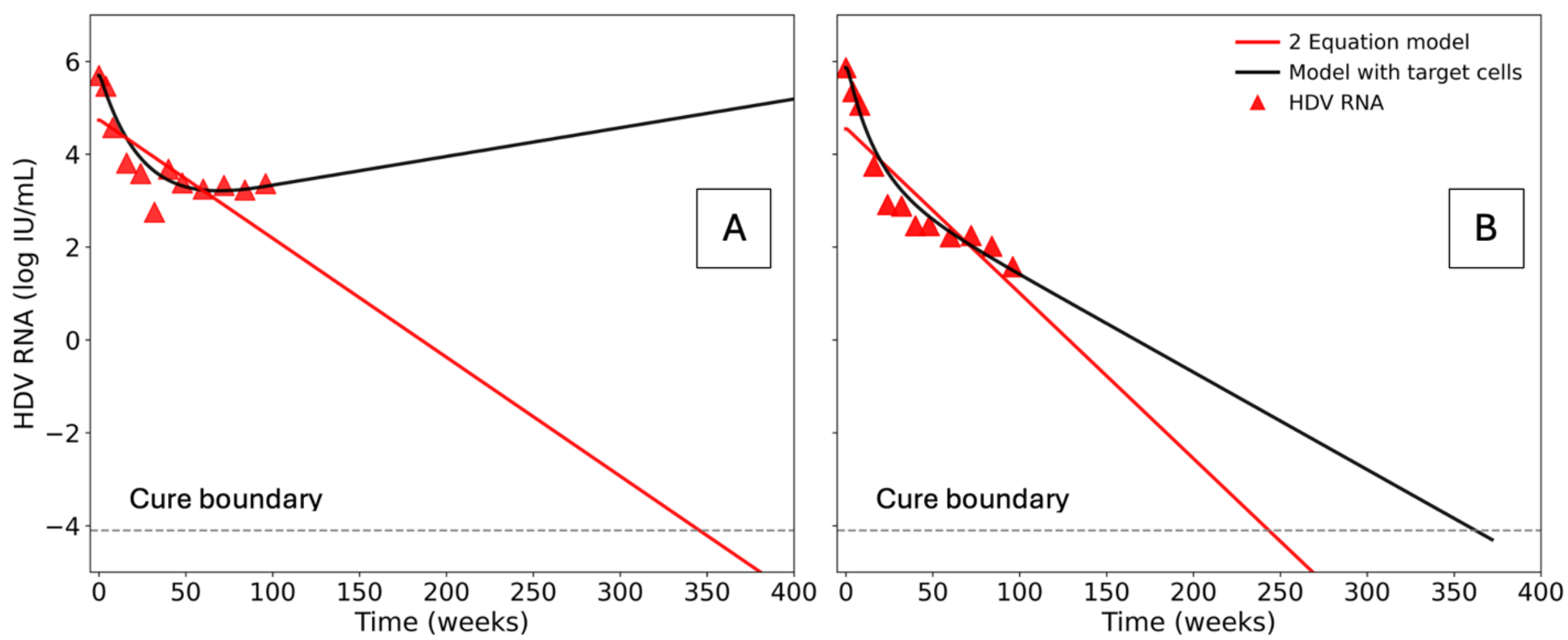

**Fig 6**. Projected time to HDV to cure for patient 3 **(A)** and patient 16 **(B)** under two models. Red line: The red line denotes the El Massoudi model and the black line represents the extended model with target cells. Triangles denote measured HDV levels.

In the El Massoudi model, the viral load remains flat at its final value before treatment cessation, which differs based on BLV efficacy as defined by $\eta$ and the length of treatment (Fig 4). In contrast, for the extended model incorporating dynamic target cells we find that the final rebound value is independent of BLV efficacy and returns to a value similar to the pretreatment level (**Fig. 7**) in accordance with recent clinical observations (Deterding et al., 2025) .

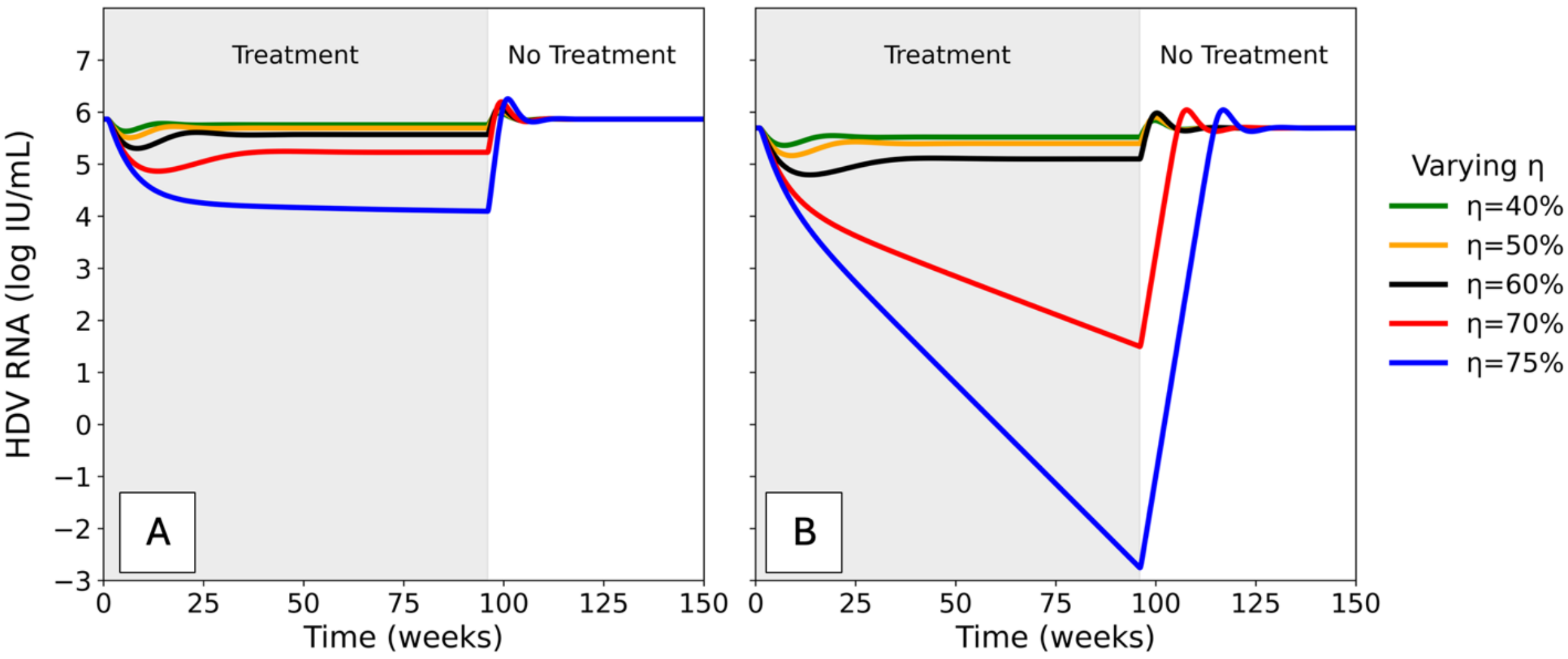

**Fig 7. Simulated treatment cessation for the model with target cells (Eq. 2); patient 3 (A) and patient 16 (B)**. The shaded grey area indicates the on-treatment phase (weeks 1 - 96); the unshaded area indicates the post treatment phase. Here we show log HDV for varying $\eta$ = [75%, 70%, 60%, 50%, 40%]. In all cases the HDV rebounds following treatment cessation.

## Conclusion

We applied the recently proposed HDV mathematical model by El Massoudi and colleagues (El Messaoudi et al., 2024) to clinical data obtained from patients treated with BLV monotherapy for up to 96 weeks using the NLME modeling approach. We showed that although the El Massoudi model parameters met the threshold for being 'precisely estimated', based on RSE < 50%, the model is structurally unable to explain many of the viral decline patterns seen during BLV monotherapy. Thus, the model cannot accurately estimate viral-host parameters , the length of BLV therapy needed to reach cure, and the viral-host dynamics after treatment is stopped. We propose an extended model including target cell dynamics, which predicts not only the monophasic viral decline pattern but also biphasic, flat-partial and viral breakthrough along with accurately predicting viral rebound once BLV is stopped.

## Supplementary Materials (SM)

### Supplementary Material Figures

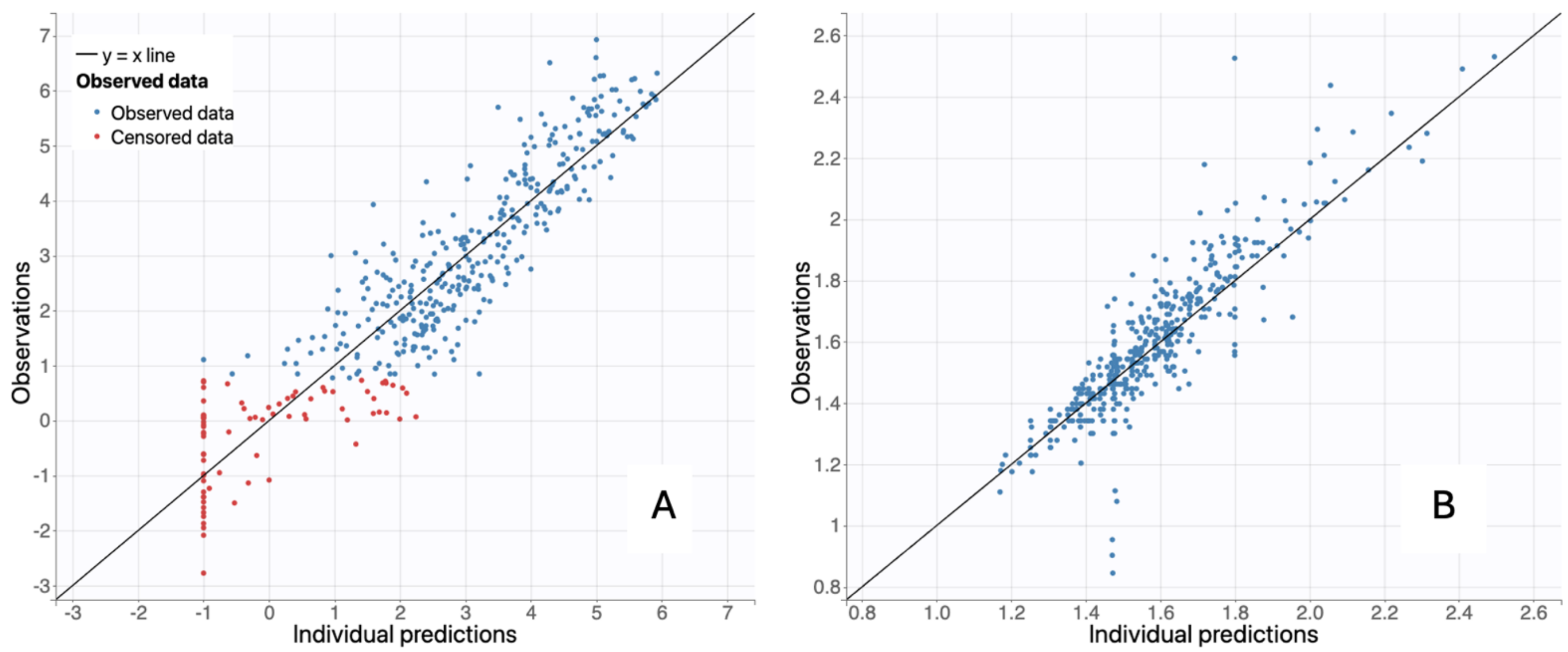

**Supplementary Fig. 1**: Plot of observations vs. individual predictions for **A** (HDV) and **B** (ALT), illustrating the relationship between observation and predictions and helping to identify potential issues with the structural model.

In **Supplementary Fig. 1A**, for HDV, the measured data points range from 4.5 to 5.5 and are not symmetrically spread around the diagonal, suggesting that the model does not accurately capture the data. Using the interactive feature in Monolix on the predictions vs. observations plots, we observed that this asymetry is because of non-monophasic declines, which were not fit well. In **Supplementary Fig. 1B**, we overall observe good agreement between the model and ALT data except in one isolated case.

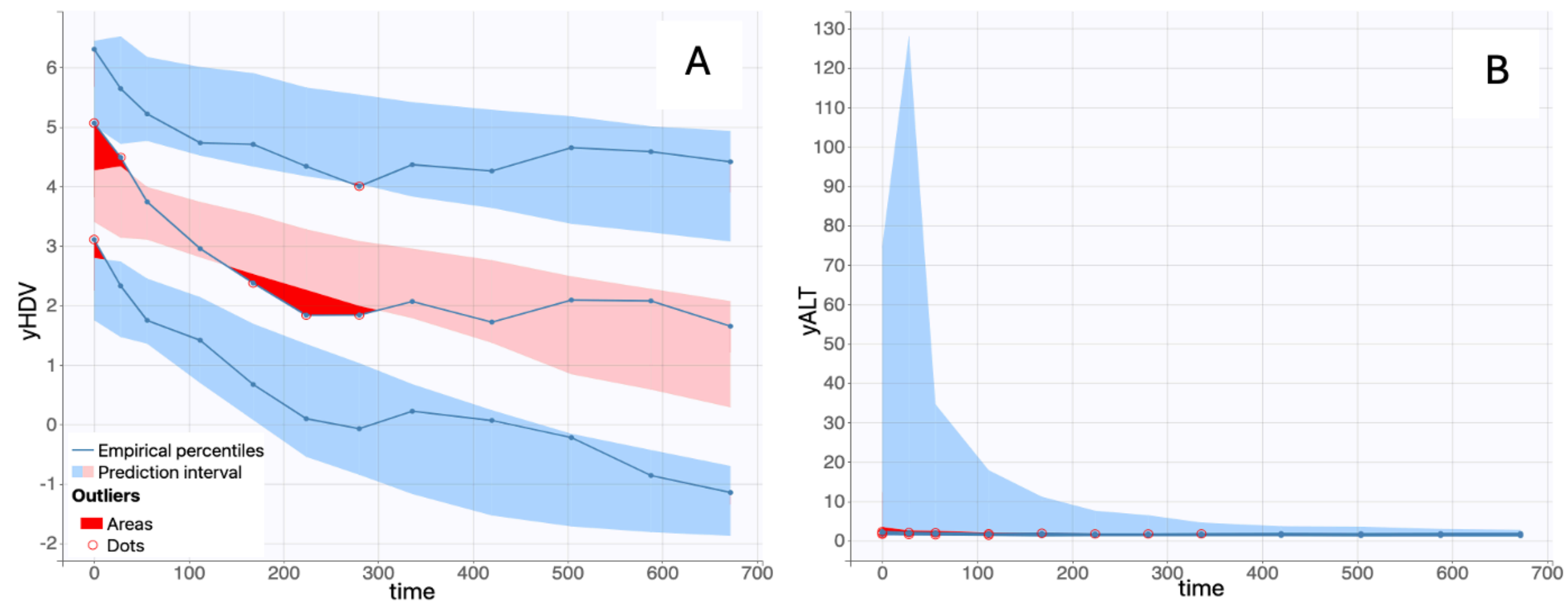

**Supplementary Fig. 2**: The ability of model (Eq. 1) to explain the data, as assesed by the visual predictive check (VPC).

The VPC is based on multiple simulations with the model and the design structure of the observed data, grouped in bins over successive time intervals with optimized binning criteria. Both Figures, **Supplementary Fig. 2A** and **2B** show that the model has good predictive power, however, its accuracy is limited by the presence of minor outliers.

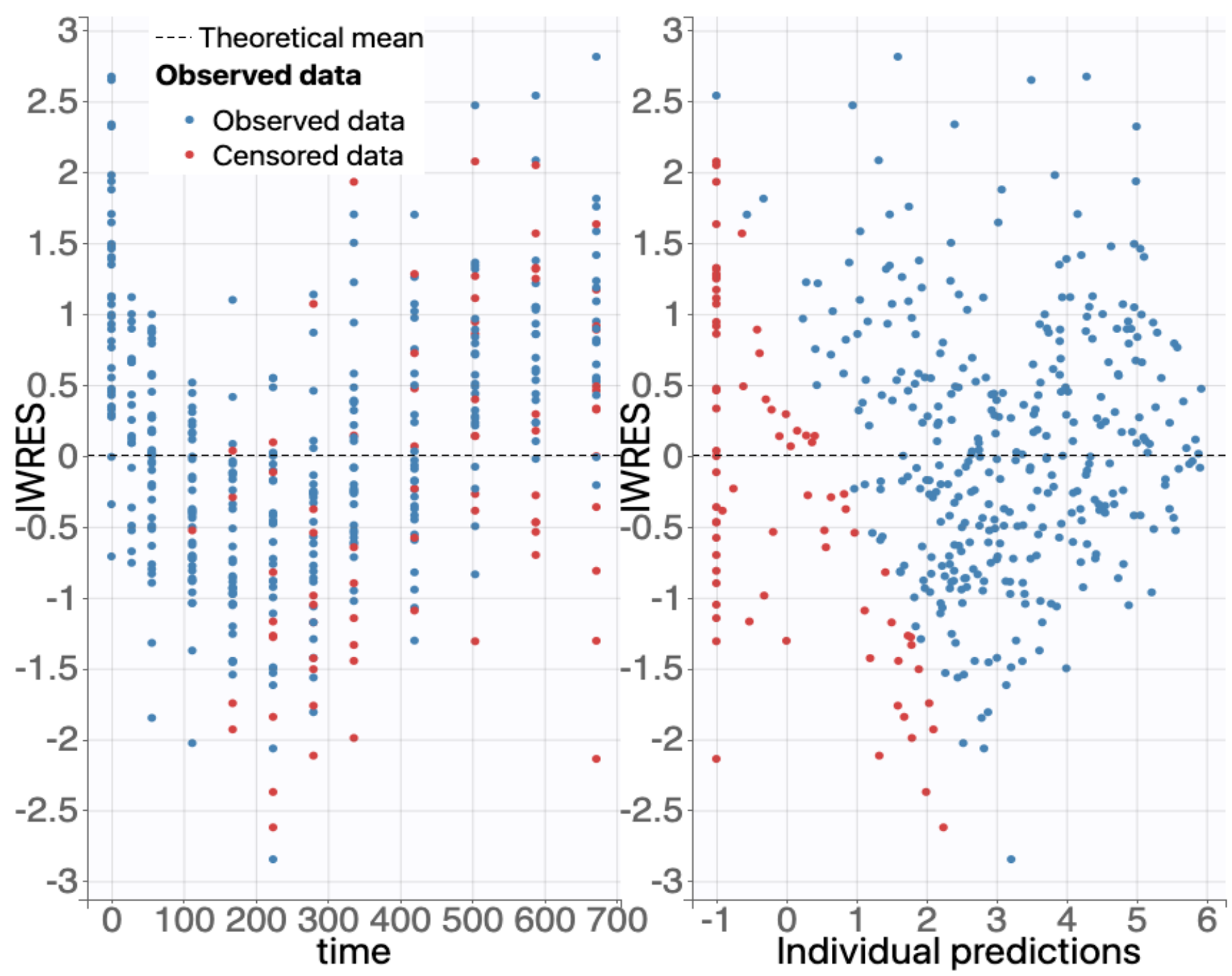

**Supplementary Fig. 3**: Diagnostic plots for model fit assessment for HDV. **(left)** Individual weighted residuals (IWRES) vs. time, denotes the disribution of individual weighted residuals over time. Ideally, the model should be randomly scattered around zero indicating an appropriate model fit with no time-related biases. **(right)** IWRES vs individual predictions, illustrating residuals in relation to individual predicted value. Ideally, residuals should be evenly distributed around zero without trends, suggesting accurate predictions across the full data set.

In **Supplementary Fig. 3**, the IWRES vs. time plot shows residuals scatterred around zero, although some clustering occurs, suggesting potential time-related model misspecifications. For the IWRES vs. individual predictions plot, the residuals are not symmetrical, indicating that the model may not fully capture variabiliry across the prediction range.

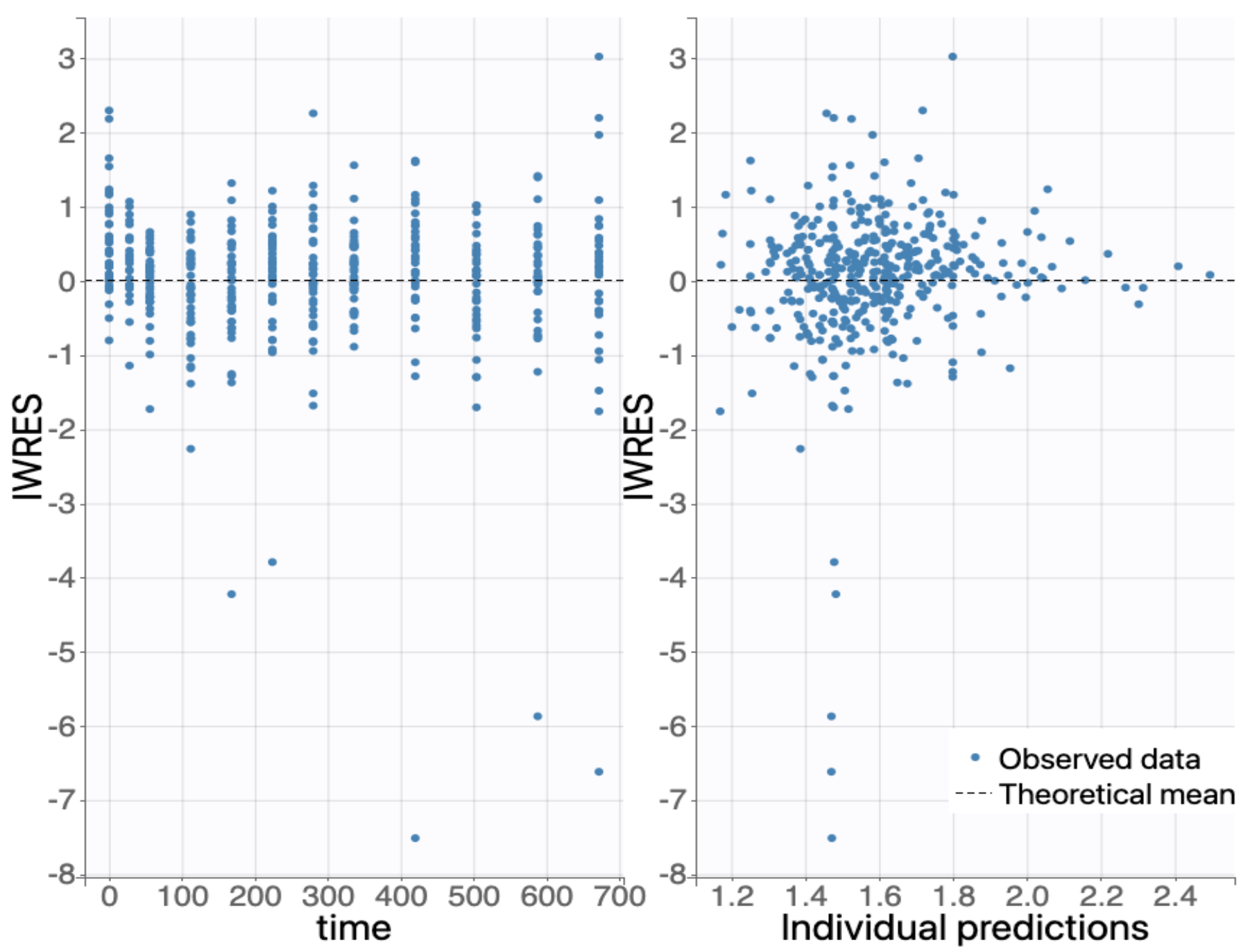

**Supplementary Fig. 4**: Diagnostic plots for model fit assessment for ALT. **(left)** IWRES vs. time, shows the disribution of individual weighted residuals over time; a random scatter around zero indicates an appropriate model fit without time-related biases. **(right)** IWRES vs. individual predictions, capturing residuals compared to individual predicted values. Ideally, residuals should be evenly distributed around zero without any patterns, suggesting accurate predictions across the range of the data.

In **Supplementary Fig. 4**, we see that in the IWRES vs. time plot, most residuals are clustered around zero, meaning that the model can explain the variability well, though a few outliers indicate potential deviations. In the IWRES vs Individual predictions plot, residuals cluster tightly around zero for most of the range, while some extreme residuals at lower and higher values indicate possible minor inaccuracie